\documentclass[aps,pra,onecolumn,superscriptaddress,notitlepage,nofootinbib]{revtex4-1}
\usepackage{mathtools}
\usepackage{amsfonts}
\usepackage{latexsym}
\usepackage{relsize}
\usepackage{mathrsfs}
\usepackage{natbib}

\DeclareMathAlphabet{\mathbbold}{U}{bbold}{m}{n}

\usepackage{euscript}%changes caligraphic font, see http://www.maths.usyd.edu.au/u/SMS/texdoc/euscript.pdf
\usepackage{amssymb}
\usepackage{graphicx}
\usepackage{amsmath}
\usepackage{amsbsy}
\usepackage{amsthm}
\usepackage{bbm}
\usepackage{bm}
\usepackage{epsfig}
\usepackage{epstopdf}
\usepackage{dsfont}

\usepackage[colorlinks]{hyperref}
\usepackage[figure,table]{hypcap}
\usepackage{enumerate}
\usepackage{geometry}
\hypersetup{
	bookmarksnumbered,
	pdfstartview={FitH},
	citecolor={darkgreen},
	linkcolor={darkred},
       linktoc={page},
	urlcolor={darkblue},
	pdfpagemode={UseOutlines}}

\usepackage{color}

\definecolor{darkgreen}{RGB}{40,130,40}
\definecolor{darkblue}{RGB}{0,0,190}
\definecolor{darkred}{RGB}{238,0,0}
\usepackage{soul}
\usepackage{float}
\usepackage{algorithm}

\floatname{algorithm}{Protocol}

\def\EQ#1{\begin{eqnarray}#1\end{eqnarray}}
\newcommand{\djj}{d\kern-0.4em\char"16\kern-0.1em}

\newtheorem{prop}{Proposition}\def\PRO{\begin{prop}}\def\ORP{\end{prop}}
\newtheorem{coro}{Corollary}\def\COR{\begin{coro}}\def\ROC{\end{coro}}
\newtheorem{theo}{Theorem}\def\TH{\begin{theo}}\def\HT{\end{theo}}
\def\TH{\begin{theo}}\def\HT{\end{theo}}
\newtheorem{defi}[prop]{Definition}\def\DE{\begin{defi}}\def\ED{\end{defi}}
\newtheorem{lemme}[prop]{Lemma}\def\LE{\begin{lemme}}\def\EL{\end{lemme}}

\def\ket#1{\left| #1 \right\rangle}
\def\bra#1{\left\langle #1 \right|}
\def\dm#1{\left|#1 \right\rangle \left\langle #1 \right|}

\usepackage{amsmath,amsfonts,amssymb}
\usepackage{wrapfig}
\usepackage{graphicx}
\usepackage{bbm}
\usepackage{graphics}

\def \beq {\begin{equation}}
\def \eeq {\end{equation}}
\def \ba {\begin{eqnarray}}
\def \ea {\end{eqnarray}}

\begin{document}

\title{Quantum mixing of Markov chains for special distributions}

\author{Vedran Dunjko}
\email{vedran.dunjko@uibk.ac.at}
%\thanks{The first two authors have contributed equally to this work.}
\affiliation{Institut f\"{u}r Theoretische Physik, Universit{\"{a}}t Innsbruck, Technikerstra{\ss}e 25, A-6020 Innsbruck, Austria}
\affiliation{Institut f\"{u}r Quantenoptik und Quanteninformation der {\"{O}}sterreichischen Akademie der Wissenschaften, Innsbruck, Austria}
\affiliation{Division of Molecular Biology, Ru\djj er Bo\v{s}kovi\'{c} Institute, Bijeni\v{c}ka cesta 54, 10002 Zagreb, Croatia.}

\author{Hans J. Briegel}

\email{hans.briegel@uibk.ac.at}
\affiliation{Institut f\"{u}r Theoretische Physik, Universit{\"{a}}t Innsbruck, Technikerstra{\ss}e 25, A-6020 Innsbruck, Austria}
%\affiliation{Institut f\"{u}r Quantenoptik und Quanteninformation der {\"{O}}sterreichischen Akademie der Wissenschaften, Innsbruck, Austria}

\begin{abstract}
%Quantum random walks offer a framework for the design of novel quantum algorithms, 
%using which polynomial, and even superpolynomial speed-ups have been reported.  %Nonetheless, even quadratic speed-ups are not guaranteed for all problems of interest.

%\red{the order on D is wrong! should be greater than}
The preparation of the stationary distribution of irreducible, time-reversible Markov chains is a fundamental building block in many heuristic approaches to algorithmically hard problems.
It has been conjectured that quantum analogs of classical mixing processes may offer a generic quadratic speed-up in realizing such stationary distributions. Such a speed-up would also imply a speed-up of a broad family of heuristic algorithms.
 However, a true quadratic speed up has thus far only been demonstrated for special classes of Markov chains.  These results often presuppose a regular structure of the underlying graph of the Markov chain, and also a regularity in the transition probabilities.
 In this work, we demonstrate a true quadratic speed-up for a class of Markov chains where the restriction is only on the form of the stationary distribution, rather than directly on the Markov chain structure itself. In particular, we show efficient mixing can be achieved when it is beforehand known that the distribution is monotonically decreasing relative to a known order on the state space. Following this, we show that our approach extends to a wider class of distributions, where only a fraction of the shape of the distribution is known to be monotonic.  Our approach is built on the Szegedy-type quantization of transition operators.
\end{abstract}

\maketitle

\section{Introduction}

Quantum walks have, amongst other reasons, been long investigated for their capacity to speed up mixing processes -- that is, speeding up the task of preparing stationary distributions of a given Markov chain (MC). Efficient mixing is a much coveted property in the context Markov Chain Monte Carlo (MCMC) approaches to many algorithmic methods for hard combinatorial problems and problems arising in statistical physics \cite{1999_Newman}. 
In the case of time-reversible Markov chains it is well-known that the bound on mixing times is tight relative to the \emph{spectral gap} $\delta$ of the Markov chain - both the lower and the upper bounds on the (approximate) mixing times are proportional to $1/\delta,$ whereas other quantities (e.g. the allowed error of the approximation) appear only as logarithmic factors.
Improvements in attaining target distributions then, in the classical case, always stem from additional constructions: e.g. by utilizing sequences of slowly evolving MCs in simulated annealing, or by using alternative MCs which mix faster toward the same distribution (e.g. graph lifting \cite{1999_Chen}). Such approaches are not proven to help generically, and their utility is established, essentially, on a case-by-case basis.

However, even without resorting to additional structures it is possible that here quantum mechanics may help generically. By employing quantum analogs of transition operators of MCs, speed-ups of mixing times already been proven \cite{2007_Richter, 2007_Richter_NJP} in the cases where the underlying transition graph corresponds to periodic lattices and the torus. In these works, the quantum operator employed was $U_t  = exp(-i D_P t),$ where $D_P$ is the discriminant operator of the (time-reversible) transition operator $P$, which is equal to $P$ itself in the cases when the stationary distribution of $P$ is uniform. The exact definition of $D_P$ will be given later. These results contribute towards a working conjecture that quantum transition operators may offer a generic quadratic speed-up in mixing times \cite{2007_Richter}.

Alternative approaches to quantum mixing, based on the Szegedy quantum transition operator, have been already proposed by Richter\footnote{The approach to quantum mixing presented here was developed before the authors were aware of the observation by Richter, and independently from the paper \cite{2007_Richter}. However, in later stages of literature review it became apparent the basic idea behind this approach was already described in Richter's paper, effectively as a side-note in the preliminaries section. }, based on observations by Childs \cite{2004_Childs, 2007_Richter}. 
In particular, it has been observed that so-called \emph{hitting} algorithms, which attempt to find a particular element in the state space, by starting from the stationary distribution of a MC, and using the transition operator, may be run in reverse to realize a \emph{mixing} algorithm.
However, to our knowledge, these approaches were not pursued further due to their inefficiency. In such an approach, the mixing time has a prohibitive dependance on the probabilities occurring in the stationary distribution, which lead an additional lower bound dependence of $\Omega(\sqrt{N})$ on the state space size $N$. Nonetheless, Szegedy-style walk operators have been successfully employed in other context, mostly relying on decreasing so-called hitting times of random walks \cite{2004_Szegedy_IEEE, 2005_Magniez,2011_Magniez_SIAM,2010_Krovi}.

In this work we re-evaluate the approach based on the Szegedy quantum transition operator (as outlined by Richter in \cite{2007_Richter}), and show how the lower bound state-space-size dependence of $\Omega(\sqrt{N})$ can be \emph{exponentially} improved to $O(\log^{3/2}(N))$ in the case when additional knowledge is available on the shape of the stationary distribution.
The outline of the paper is as follows: in section \ref{Sect3} we give the preliminaries and set up the notation. Following this, in section \ref{Sect4} we explain the main result for a special case of monotonically decaying distributions. In section \ref{sect5} we explain when and how the result can be extended to a much wider class of distributions, elaborate why similar techniques cannot be useful in classical mixing problems, and also prove the optimality of our approach.
We finish off in section \ref{Sect6} with a brief discussion. 

\section{Preliminaries}
\label{Sect3}
We begin by a brief recap of the basic concepts and results in discrete-time, time-homogeneous Markov chain theory. 
A discrete-time, time-homogeneous Markov chain is characterized 
with a transition matrix (operator) $P$ which acts on a state space $S$ of $N$ states. P can be represented by a left-stochastic matrix (a matrix with non-negative, real entries with columns adding to one). $P,$  with an initial distribution, fully specify a Markov chain and we will often refer to $P$ as the transition matrix and the Markov chain, interchangeably. If $P$ is irreducible (that is, $P$ is an adjacency matrix of a strongly connected graph) and aperiodic (the greatest common divisor of the periods of all states is 1), then there exists a unique stationary distribution $\pi,$ such that $P \pi = \pi$ \footnote{In this work the transition matrices are left-stochastic and act on column-vectors from the left.}.
We will represent distributions as a non-negative column vector $\pi = (\pi_i)_{i=1}^{N},$ $\pi_i \in \mathbbmss{R}^{+}_0$, such that $\sum_i \pi_i =1.$ 
Irreducible and aperiodic MCs mix: a sequential application of $P$ onto any initial state in the limit yields the stationary distribution. More precisely, it holds that $\lim_{t\rightarrow \infty}P^t \sigma = \pi,$ for all initial distributions $\sigma.$ This property is sometimes referred to as \emph{the fundamental theorem of Markov chains}.

In this work we will focus on time-reversible, irreducible and aperiodic Markov chains. A Markov chain $P$ with stationary distribution $\pi$ is said to be time-reversible if it satisfies detailed balance:
\EQ{
\pi_i P_{ij} = \pi_j P_{ji}, \forall\ i,j.
}
More generally, for an irreducible, aperiodic Markov chain $P,$ over the state space of size $N$ with stationary distribution $\pi$, we define the time-reversed Markov chain $P^{\ast}$ with $P^{\ast} = M(\pi) P^{T} M(\pi)^{-1},$ where $M$ is the diagonal matrix $M=diag(\pi_1, \ldots, \pi_N).$\footnote{The inverse of $D$ always exists, as stationary distributions of irreducible aperiodic Markov chains have non-zero support over the entire state space.} Then, $P$ is time-reversible if $P=P^{\ast}.$ The discriminant matrix $D_P$ is then defined with $D_P = M^{1/2}(\pi) P^{T} M(\pi)^{-1/2},$ and it can be shown that it is always symmetric for time-reversible Markov chains. 
Since a time-reversible transition matrix $P$ is similar to a symmetric matrix, its eigenvalues are real, and also by the Perron-Frobenius theorem they are less or equal to $1$ in absolute value (value 1 being reserved for the stationary distribution which is also the +1 eigenvector).

If $\lambda_2$ denotes the second largest eigenvalue (in absolute value) then with $\delta = 1 - |\lambda_2|$ we will denote the \emph{spectral gap} of the Markov chain $P$.

Next, the mixing time $\tau(\epsilon),$ within error $\epsilon$, for $P$ is defined as:
\EQ{
\tau(\epsilon)  = \min \left\{t \vert D(P^{t} \sigma,\pi)\leq \epsilon, \forall \sigma \right\}
}
where $D(\pi, \sigma)$ denotes the total variational distance on distributions $\pi, \sigma$, so $D(\pi, \sigma) = 1/2 \sum_j |\pi_j  - \sigma_j|$, or, more generally, the trace distance on density matrices $\hat{\pi}, \hat{\sigma}$: $D(\hat{\pi}, \hat{\sigma}) = 1/2\, Tr[|\hat{\pi}- \hat{\sigma}|]$.

The mixing time (for the MC $P$, with stationary distribution $\pi$) has a tight bound, in the time-reversible case, proven by Aldous in \cite{1982_Aldous}, but which we present in a more detailed form derived from \cite{2006_Levin}:

\EQ{
\lambda_2 \dfrac{1}{\delta} \log\dfrac{1}{2 \epsilon} \leq \tau(\epsilon) \leq  \dfrac{1}{\delta} (\log\dfrac{1}{\pi_{min}} + \log\dfrac{1}{\epsilon}) \label{mixBounds}
}
for $\pi_{min} = \min_{i}\, \pi_i$.

 For more details on Markov chains, we refer the reader to  \cite{1998_Norris}.

Next, we present the basic elements of Szegedy-style approaches to quantum walks.
Part of the presentation follows the approach given in \cite{2011_Magniez_SIAM}.

Szegedy-style quantum walks can be viewed as walks over a bipartite graph, realized by duplicating the graph of the original MC, specified by the transition matrix P.
The basic building block is a diffusion operator $U_P$ \footnote{While there are infinitely many such operators, any one of them will serve our purpose.} which acts on two quantum registers of $N$ states and satisfies 
\EQ{
U_{P} \ket{i}_{\textup{I}}\ket{0}_{\textup{II}} = \ket{i}_{\textup{I}} \sum_{j=0}^{N-1}\sqrt{P_{ji}} \ket{j}_{\textup{II}}.\label{UP}
}

It is easy to see that $U_P$ establishes a walk step from the first copy of the original graph to the second.
The operator $U_P,$ and its adjoint are then used to construct the following operator:
\EQ{
ref(A) = U_P (\mathbbmss{1}_{\textup{I}} \otimes Z_{\textup{II}}) U_P^{\dagger},
} 
where $Z = 2 \ket{0}\bra{0} - \mathbbmss{1},$ reflects over the state $\ket{0}$.
The operator $ref(A)$ is a reflector  the subspace $A = \textup{span}(\{ U_P \ket{i} \ket{0} \}_i)$.
A second reflector is established by defining a second diffusion operator, realizing a walk step from the second copy of the graph back to the first: $V_P = SWAP_{\textup{I},\textup{II}} U_P SWAP_{\textup{I},\textup{II}}.$ From here, we proceed analogously as in the case for the set $A$, to generate the $ref(B)$ operator, reflecting over $B = \textup{span}(\{ V_P \ket{0} \ket{j} \}_j)$.
The Szegedy walk operator is then defined as $W(P) = ref(B) ref(A)$. 
In \cite{2004_Szegedy_IEEE,2011_Magniez_SIAM} it was shown that the operator $W(P)$ and $P$ are closely related, in particular in the case $P$ is time-reversible, which we clarify next.

Given a distribution $\pi,$ we will denote the \emph{coherent encoding of the distribution $\pi$} with $\ket{\pi} = \sum_{i=1}^{N} \sqrt{\pi_i} \ket{i}$.
For us it is convenient to define a one-step diffused version of the encoding above, specific to a particular Markov chain: $\ket{\pi'} =U_{P} \ket{\pi}_{\textup{I}} \otimes \ket{0}_{\textup{II}}$, where $U_P$ is the Szegedy diffusion operator.
It is easy to see that $\ket{\pi}$ and $\ket{\pi'}$ are trivially related via the diffusion map (more precisely, the isometry $\ket{\pi} \rightarrow U_P \ket{\pi} \otimes \ket{0}$) and moreover that the computational measurement of the first register of $\ket{\pi'}$ also recovers the distribution $\pi$.
By abuse of notation, we shall refer to both encodings as \emph{the coherent encoding} of the distribution $\pi,$ and denote them both $\ket{\pi},$ where the particular encoding will be clear from context.  
Next, we clarify the relationship between the classical transition operator $P$ and the Szegedy walk operator $W(P).$ Let $\pi$ be the stationary distribution of $P$ so $P \pi = \pi$. 
 Then the coherent encoding of the stationary distribution $\pi$ of $P$, given with $\ket{\pi} = U_P \sum_{i} \sqrt{\pi_i} \ket{i} \ket{0},$ is also a +1 eigenstate of $W(P)$, so $W(P)\ket{\pi} = \ket{\pi}$. Moreover, on the subspace $A+B$, so-called \emph{busy subspace}, it is the unique $+1$ eigenstate. On the orthogonal complement of the busy subspace, $W(P)$ acts as the identity.
 Moreover, the spectrum of $P$ and $W(P)$ is intimately related, and in particular the spectral gap $\delta$  is quadratically smaller than the phase gap \EQ{\Delta = \min\left\{  2 | \theta| | e^{i \theta} \in \sigma\left(W\left(P\right)\right), \theta \not= 0\right\} , }  where $\theta$ denote the arguments of the eigenvalues, i.e. eigenphases, of $W(P)$.
In other words, we have that $1/\Delta \in O(1/\sqrt{\delta}).$
 This relationship is at the very basis of all speed up which stems from employing Szegedy-style quantum walks.
We refer the reader to \cite{2011_Magniez_SIAM,2004_Szegedy_IEEE} for further details on Szegedy-style quantum walks.

A useful central tool in the theory of Szegedy-style quantum walks is the so-called Approximate Reflection Operator $ARO(P) \approx 2\dm{\pi} - \mathbbmss{1}$, which approximately reflects over the state $\ket{\pi}$.
The basic idea for the construction is as follows:
By applying Kitaev's phase detection algorithm on $W(P)$ (with precision $O(\log(\Delta))$), applying a phase flip to all states with phase different from zero, and by undoing the phase detection algorithm, we obtain an arbitrary good approximation of the reflection operator $R(P) = 2 \dm{\pi} - \mathbbmss{1}$, for any state within $A+B$.
The errors of the approximation can again be efficiently suppressed by iteration (by the same arguments as for the $\ket{\pi}$ measurement) \cite{2011_Magniez_SIAM}, so the cost of the approximate reflection operator is in $\tilde{O}(1/\Delta) = \tilde{O}(1/\sqrt{\delta}).$ 

Thus, the second gadget in our toolbox is the operator $ARO(P),$ which approximates a perfect reflection $R(P)$ on $A+B$, while incurring a cost of $\tilde{O}(1/\sqrt{\delta})$ calls to the walk operator $W(P)$.

The  $ARO(P)$, along with the capacity to flip the phase of a chosen subset of the computational basis elements, suffices for the implementation of an amplitude amplification \cite{2000_Brassard} algorithm which allows us to find the chosen elements. 
To illustrate this, assume we are given the state $\ket{\pi},$ the (ideal) reflector $R(P),$ and assume we are interested in finding some set of elements $M \subseteq \{1, \ldots, N \}$.
The subset $M$ is typically specified by an oracular access to a phase flip operator defined with$Z_M = \mathbbmss{1} - 2\sum_{i \in S} \dm{i}$.
The element searching then reduces to iterated applications of $Z_M R(P)$ (which can be understood as a generalized Grover iteration, more precisely amplitude amplification) onto the initial state $\ket{\pi}.$ 
Let $\tilde{\pi}$ denote the conditional probability distribution obtained by post-selecting on elements being in $M$ from $\pi,$ so
\EQ{
\tilde{\pi} = \left\lbrace \begin{tabular}{cl}\vspace{0.1cm}
$\dfrac{\pi_i}{\epsilon},$& $\textup{if} \ i\in M$\\
0,& \textup{otherwise},
\end{tabular}  \right. \label{EQ3}
} with $\epsilon = \sum_{j \in M} \pi_j.$ Let $\ket{\tilde{\pi}} = U_P \sum_i \sqrt{\tilde{\pi}_i} \ket{i} \ket{0}$ denote the coherent encoding of $\tilde{\pi}.$  Note that the measurement of the first register of $\ket{\tilde{\pi}}$ outputs an element in $M$ with probability 1. Thus successfully preparing this state implies that we have found a desired element from $M$. 

As it was shown in \cite{2011_Magniez_SIAM}, applications of $Z_M $, and $R(P)$ maintain the register state in the two-dimensional subspace $\textup{span}( \{ \ket{\pi}, \ket{\tilde{\pi}} \}),$ and moreover, using $\tilde{O}(1/\sqrt{\epsilon})$ applications of the two reflections will suffice to produce a state $\ket{\psi} \in span  \{ \ket{\pi}, \ket{\tilde{\pi}} \},$ such that $| \bra{\psi} \tilde{\pi} \rangle |^2$  is a large constant (say above 1/4). Measuring the first register of such a state will result in an element in $M$ with a constant probability, which means that by iterating this process $k$ times ensures an element in $M$ is found with an exponentially increasing probability in $k$.
However, since the state $\ket{\psi}$ is also in $ span  \{ \ket{\pi}, \ket{\tilde{\pi}} \},$ it is easy to see that the measured state, conditional on being in $M,$ will also be distributed according to $\tilde{\pi}$. This observation was used in \cite{2014_Paparo}, and also in \cite{2010_Krovi} to produce an element sampled from the truncated stationary distribution $\tilde{\pi},$ in time $\tilde{O}(1/\sqrt{\epsilon})\times \tilde{O}(1/\sqrt{\delta})$ where the $\delta$ term stems from the cost of generating the approximate reflection operator $ARO(P)$, and $\tilde{O}(1/\sqrt{\delta})$ corresponds to the number of iterations which have to be applied. This is a quadratic improvement relative to using classical mixing, and position checking processes which would result in the same distribution.

However, the same process can be used \emph{in reverse} to generate the state $\ket{\pi}$ starting from some fixed basis state $\ket{i'} = U_{P} \ket{i} \ket{0} $ 
with cost $\tilde{O}(1/\sqrt{\delta}) \times \tilde{O}(1/\sqrt{\pi_i})$.  The resulting state of the reverse process is constantly close to the state $\ket{\pi},$ which is our target state. This basic idea was already observed in \cite{2007_Richter} by Richter, however, at first glance it seems prohibitive as the resulting mixing time is proportional to $\tilde{O}(1/\sqrt{\pi_i})$. If no assumptions are made on the stationary distribution, this dependency is lower bounded by $\tilde{O}(1/\sqrt{N}),$ as the smallest probability in a distribution is upper bounded by $1/N$.

For this work, we point out that the preparation process above, which starts from an initial basis state is trivially generalized:  if $\ket{\psi}$ is any initial state, and we have the capacity to reflect over it,  we can reach a state close to the target state $\ket{\pi},$ with an overall cost $\tilde{O}(1/\sqrt{\delta}) \times O(1/\sqrt{F(\ket{\psi}, \ket{\pi})})$, where $F(\ket{\psi}, \ket{\pi})= |\bra{\psi} \pi \rangle |^2$ is the standard fidelity. This follows as the search/unsearch algorithms are in fact amplitude amplification \cite{2000_Brassard} algorithms.

Finally we point out that having a constant-distance approximation of the stationary distribution is effectively all we need. Given an approximation, which is constantly far from $\ket{\pi}$, an arbitrarily good approximation of distance $\epsilon$ can be achieved in time $O(1/\sqrt{\delta} \times \log(1/\epsilon))$, by again running phase estimation (iteratively) of $W(P)$ on the approximate state, and this time measuring the phase-containing register. This $\ket{\pi}$-projective measurement is described in more detail in \cite{2015_Dunjko}, and it follows from Theorem 6 in \cite{2011_Magniez_SIAM} .

We have previously used this idea in \cite{2015_Dunjko}, to achieve more efficient mixings in the context of slowly evolving sequences of Markov chains, where the initial states were either basis states, or a state encoding the uniform distribution.

In this work, we will take this idea significantly further, by intrinsic properties of monotonically decaying distributions. Let $\Omega$ be a finite state space, and let $\leq_{\Omega}$ be a total order on  $\Omega$. Then a distribution $d$ is monotonically decaying, relative to the order $\leq_{\Omega}$, if for its probability mass function $f_{d}$ we have: $x,y \in \Omega,\ (x \leq_{\Omega} y) \Rightarrow (f_{d}(x) \geq f_{d}(y)).$ 
In this work we will be representing the state space elements with integers, and the order will be the standard order so a distribution $\pi$ is monotonically decaying if $i\leq j \Rightarrow \pi_i \geq \pi_j$, monotonically increasing if $i\leq j \Rightarrow \pi_i \leq \pi_j$. Finally, a distribution $\pi$ is strongly unimodal if there exists a $k \in \Omega$ such that for $i \leq j \leq k$ we have $\pi_i \leq \pi_j$ and for  $k \leq i \leq j $ we have $\pi_i \geq \pi_j$.
\section{Main result}
\label{Sect4}
The results of the previous section already establish that if we have an (perhaps oracular) access to good approximations of the targeted stationary distribution, arbitrarily good mixing by unsearching is efficient.
Here, we will show that, for the case of monotonically decaying distributions we can construct good initial states efficiently, in time $O(\log(N))$, independently from the shape of the distribution.
Our first result is given with the following theorem:
\TH \label{first}
Let $N = 2^n, n \in \mathbbmss{N}$ be an integer, and let $D^{\geq}_{N} \subseteq \mathbbmss{R}^{N}$ be the convex space of monotonically decaying distributions over $\{1, \ldots, N \}.$
Then there exists a set of distributions $S \subseteq D^{\geq}_{N}$, $|S| = n,$ such that for every $\pi \in D^{\geq}_{N},$ there exists $\nu \in S$ satisfying 
\EQ{
D(\pi, \nu) \leq 1- \dfrac{1}{2(n+1)}.
}
\HT
 \proof
 We begin by constructing an $N$-sized set $S' = \{\sigma^i \}_{i=1}^{N}$ of "ladder" distributions which are extreme points of the convex space $D^{\geq}_{N}$. They are defined as:
 \EQ{
 (\sigma^i)_j =   \left\lbrace \begin{tabular}{cl}\vspace{0.1cm}
$1/i$& $\textup{if} \ j \leq i$\\
0,& \textup{otherwise}.
\end{tabular}  \right. \label{ladders}
 } 
Any distribution $\pi \in D^{\geq}_{N}$ can be represented as a convex combination of distributions in $S'$ as follows:
\EQ{
\pi = \pi_1 \sigma^1 + \sum_{i=2}^{N} \pi_i (i\sigma^i - (i-1)\sigma^{i-1}),
}
as the parentheses above contain the $i^{th}$ Kronecker-delta distribution.
The expression above can be, by reshuffling, restated as
\EQ{
\pi = \left(\sum_{i=1}^{N-1} i (\pi_i - \pi_{i+1}) \sigma^i \right) + N \pi_N \sigma^{N} = \sum_{i=1}^{N}  q_i \sigma^i, \label{eq1}
 }
 where, for $1 \leq i < N$ $q_i = i  (\pi_i - \pi_{i+1})$ and $q_N = N \pi_N$. Since the distribution $\pi$ is monotonically decaying, we have that $q_i \geq 0,$ and it is also easy to see that $\sum_i q_i = 1.$ In other words, $q = (q_i)_i$ is a distribution as well.
Using the representation above, we can express the distance between $\pi$ and $\sigma^k \in S'$ as follows:
\EQ{
D(\pi, \sigma^k) = D( \sum_{i=1}^{N}  q_i \sigma^i, \sigma^k) \leq \sum_{i=1}^{N} q_i D(\sigma^i, \sigma^k),
}
where the last inequality follows from the strong convexity of the trace distance, and in fact is a strict equality in our case. The distance between individual distributions in $S'$ is easy to express explicitly:
\EQ{
D(\sigma^i, \sigma^k) = 1 - \dfrac{\min\{i,k \}}{\max\{i,k \}}.
}

If we define the matrix $V$ as $V_{ij} = \dfrac{\min\{i,k \}}{\max\{i,k \}}$, we can express the trace distance between $\pi$ and $\sigma^k$ as:
\EQ{
D(\pi, \sigma^k) = 1- v_k^T  q, 
}
that is 1 minus the standard inner product between the $k^{th}$ column of $V$, denoted $v_k$, and the probability vector $q$ which uniquely specifies $\pi$.
Next, we focus on the log-sized subset $S \subseteq S'$ given with $S = \{\sigma^i  \vert i = 2^k, k = 0,\ldots, n-1\},$ and establish a lower bound on $\min_q \max_k (v_{2^k})^T  q$ by coarse-graining. This will then yield an upper bound on the distance between an arbitrary decaying distribution $\pi$ and the set $S$.

From the definition of the vectors $v_{2^k}$, $k<n$ it is easy to see that the following holds:
\EQ{
(v_{2^k})_j \geq 1/2\  \textup{for\ all}\ 2^k \leq j \leq 2^{k+1}. \label{eq3}
}
We can see this more generally as, per definition, for $v_l,$\ $l\leq N/2 = 2^{n-1}$ and $l\leq j \leq 2 l$ we have that $(v_l)_j = \dfrac{l}{j} \geq \dfrac{l}{2 l} = 1/2 .$ 
Next, we introduce the coarse-graining operator $\mathcal{L}$, mapping real vectors over $2^n$ elements to real vectors over $n+1$ elements:

\EQ{
\mathcal{L}(q) = (\tilde{q}_j)_{j=1}^{n+1} \ \emph{with}\\
\tilde{q}_1 = q_1, \textup{and\ for}\ j>1, \tilde{q}_j = \sum_{l =2^{j-2}+1}^{2^{j-1}} q_l.
}

It is clear that $\mathcal{L}$ also maps distributions to coarse-grained distributions.
By Eq. (\ref{eq3}) we have that
\EQ{
v_{2^k} \geq w_{2^k}, \textup{for}\ k<n,\  \textup{with}\\
 (w_{2^k})_j = \left\lbrace \begin{tabular}{cl}\vspace{0.1cm}
$1/2$& $\textup{if} \ 2^k \leq j \leq 2^{k+1}$\\
0,& \textup{otherwise},
\end{tabular}  \right. 
}
where the inequality is taken element-wise. The ancillary vectors $w_{2^k}$ just capture the positions where the vector $v_{2^k}$ has entries larger or equal to 1/2, setting those to 1/2 and the rest to zero.
But then it follows, for $0\leq k \leq n-1$, that
\EQ{
v_{2^k}^T  q \geq w_{2^k}^T  q,
}
where the inequality is taken element-wise.
Next, with $\Delta_k$ we denote the Kronecker-delta distribution with unit support only over the $k^{th}$ element, and the inequalities are element-wise, on the coarse-grained $n+1$ element space.
It holds that
\EQ{
(w_{2^0}^T ) q \geq \dfrac{1}{2} \Delta_{k} \cdot \mathcal{L}(q), \textup{for }\ 1\leq k \leq 2,\ \textup{and}\\
(w_{2^k}^T)  q \geq \dfrac{1}{2} \Delta_{k+2} \cdot \mathcal{L}(q), \textup{for }\ 0< k<n.
}
To see the first claim, note that $(w_{2^0}^T )q = 1/2 (q_1 + q_2),$ whereas $\Delta_{1} \cdot \mathcal{L}(q) = q_1$ and $\Delta_{2} \cdot \mathcal{L}(q) = q_2$.
For the second inequality, note that the left hand side of the inequality sums up all elements from $q$ which lie between positions $2^k$ and  $2^{k+1}$ (boarder points included), and multiplies it with 1/2. The right hand side of the inequality picks out the $(k+2)^{nd}$ element of the coarse-grained distribution $\mathcal{L}(q),$ which, by definition, sums the entries of $q$ between the same boundaries, but not including the lower boundary.
Then we have that 
\EQ{
\max_{k \in \{1,\ldots, n+1\}}\ v_{2^{k-1}}^T  q \geq \max_{k \in \{1,\ldots, n+1\}}\ w_{2^{k-1}}^T  q \geq  \dfrac{1}{2 }\max_{k \in \{1,\ldots, n+1\}}\ \Delta_k \cdot  \mathcal{L}(q). 
}

%By definition we have that the (scaled) coarse-grained vectors $w$ upper-bound the extremal distributions of the $n+1$ dimensional space:
%\EQ{
%\mathcal{L}(w_{2^0}) \geq 1/2 \Delta_0, \textup{and}\  \textup{for} \ 1\leq k \leq n, \\
%\mathcal{L}(w_{2^{k-1}}) \geq 1/2 \Delta_k,
%}
%where $\Delta_k$ is the Kronecker-delta distribution with unit support only over the $k^{th}$ element, and the inequalities are element-wise.
%
%Since all the non-zero entries in vectors $w_{2^k}$ have the same value 1/2, we have that
%\EQ{
%(w_{2^0})^T . q = \mathcal{L}(w_{2^0}). \mathcal{L}(q) \geq 1/2 (\Delta_0)^{T}. \mathcal{L}(q)\ \textup{and}\  \textup{for} \ 1\leq k \leq n, \\
%(w_{2^{k-1}})^T . q  = \mathcal{L}(\ora{w}_{2^{k-1}}). \mathcal{L}(q) \geq 1/2 (\Delta_k)^{T}. \mathcal{L}(q)
%}
Then, the target min-max expression is also lower bounded by  \EQ{ \min_{q} \max_{k \in \{1,\ldots, n+1\}} \dfrac{1}{2} \Delta_k \cdot \mathcal{L}(q),} which is easy to evaluate: $\mathcal{L}(q)$ is an arbitrary distribution over $n+1$ elements, and we are free to optimize the overlap of this distribution with all Kronecker-delta distributions on the same space. The minimum is attained when all the overlaps are the same, so when $\mathcal{L}(q)$ is uniform over the space of $n+1$ elements, and we have $\ \min_{q} \max_{k \in \{1,\ldots, n+1\}} 1/2 \Delta_k^{T} \cdot \mathcal{L}(q) = \dfrac{1}{2(n+1)}$.
This also lower bounds our target expression  $\min_q \max_k  (v_{2^k})^T . q$, and proves our claim. \qed

In the proof above we have explicitly constructed the $\log(N)$ distributions from the set $S$. They are the $n = \log_2(N)$ distributions $\nu^k \mathrel{\mathop:}= \sigma^{2^k}$, for $0\leq k \leq n-1$  which have uniform support from the first element up to element at the $(2^k)^{th}$ position.
For them to be useful for the quantum mixing algorithm the coherent encodings of these distributions have to have an efficient construction, which is the case. 
Start by initializing the $n-$qubit register (sufficient for encoding distributions over the $N=2^n$ state-space) in the "all-zero" state. Then, the $k^{th}$ distribution is achieved by applying the Hadamard gate to the last $k$ qubits. This realizes the state $\ket{\nu^k} = \ket{0}^{\otimes(n-k)} \ket{+}^{\otimes(k)},$ which encodes the desired distributions, and the reverse of this process, along with the reflection over the "all zero" state realizes the reflection over $\ket{\nu^k}$ efficiently as well.

A few remarks are in order. First, although we have phrased the result for the case when the state space is a power of 2, this is without loss of generality -- any decaying distribution over N elements is trivially a decaying distribution over the larger set of $\lceil \log_2(N) \rceil$ elements, where we assign zero probability to the tail of the distribution. The trace distance result remains the same, once the ceiling function is applied to the log term, hence yields the same scaling.

Next, note that the Theorem \ref{first}, along with the given simple method for preparing the log-sized set of initial states already yields an efficient algorithm for the preparation of decaying stationary distributions. 
To see this, assume first we know which distribution $\nu^k$ out of $S$ minimizes the trace distance, bounded by $1-\dfrac{1}{2(\log(N)+1)}.$
If $\ket{\pi}$ is the coherent encoding of the target stationary distribution, by the known inequalities between the trace distance and the fidelity\footnote{Note that the total variation distance on the distributions, corresponds to the trace distance of the incoherent encodings of probability distributions, whereas we are interested in the fidelity of the coherent encodings. However, the Uhlmann fidelities of coherent and incoherent encodings are equal, so the standard bounds do apply.}  we have that:
\EQ{
|\bra{\pi} \nu^k \rangle|^2 \geq \dfrac{1}{(2(\log(N)+1))^2}.
} 
But then, by the results of section \ref{Sect3}, we can attain the stationary distribution in time $\tilde{O}(1/\sqrt{\delta}) \times O(2(\log(N)+1))$ $=\tilde{O}(1/\sqrt{\delta}) \times O(\log(N)),$ where the soft-O ($\tilde{O}$) part suppresses the logarithmically contributing factor stemming from the acceptable error term.
However, since we do not know which distribution minimizes the distance, the trivial solution is to sequentially run the algorithm for each one. This yields an overall $\log(N)$ factor, yielding  $\tilde{O}(1/\sqrt{\delta}) \times O(\log^{2}(N))$ as the total complexity.
We can do slightly better by encasing the entire procedure of "searching for the correct initial distribution" in a Grover-like search algorithm, more precisely, an amplitude amplification prodecure. 

To see this is possible, note that whether or not a particular distribution was the correct initial choice can be heralded - the $\ket{\pi}-$projective measurement, for instance, reveals whether we succeeded or did not. Moreover, the $ARO(P)$ operator itself will help realize the Grover oracle, which flips the phase of all states which are not the target distribution.
The overall procedure can then be given as follows:

First, we initialize the system in the state $\sum_{j=0}^{n-1} \ket{j}_{\textup{I}} \ket{\psi_j}_{\textup{II}}$, where $\ket{\psi_j}$ is the coherent encoding of the $j^{th}$ distribution from the set $S$. This has a complexity of $O(\log^{2}(N))$ in the state-space size, but is independent from $\delta$. Then, in quantum parallel, we run the quantum mixing algorithm on register $\textup{II}$ (with complexity $\tilde{O}(1/\sqrt{\delta}) \times O(\log(N)))$, followed by one application of the $ARO(P)$, followed by an un-mixing (the running of the mixing algorithm in reverse). This will, approximately (and up to a global phase of $-1$), introduce a relative phase of $-1$ at those $\ket{j}$ terms, where the searching procedure yielded a success. This constructs the phase-flip operator.

 The remainder is the operator which flips over the state $\sum_{j=0}^{n-1} \ket{j}_{\textup{I}} \ket{\psi_j}_{\textup{II}}$, which has a cost of $O(\log^{2}(N))$.
Since at least one distribution, by the correctness of our mixing algorithm, yields the target distribution, this extra layer of amplitude amplification needs to be run in a randomized fashion, (since only the lower bound is known) \cite{2000_Brassard}, on the order of $\sqrt{\log(N)}$ times, until the correct initial distribution is found.
The overall complexity, is then given with $O\left(1/\sqrt{\delta} \times \log^{3/2}(N) \right) + O\left( \log^{5/2}(N) \right).$ The the error factor (multiplying both additive terms) which we have for simplicity omitted, and which guarantees that the distance from the target distribution is within $\epsilon$ in the trace distance, is given with $O(\log(1/\epsilon) + \log \log(N) ).$ The additional $\log \log(N) $ term stems from the fact that the $ARO(P)$ operator is applied $O(\log(N))$ many times which accumulates errors. However, since the effective total error is given by the union bound, it will suffice to rescale the target precision by $\epsilon := \epsilon/\log(N),$ which yields the $\log\log$ term \cite{2011_Magniez_SIAM}. In practice, $1/\sqrt{\delta}$ tends to dominate $\log(N),$ thus we have the complexity  $O\left(1/\sqrt{\delta} \times \log^{3/2}(N) \right)$, omitting the logarithmically contributing error terms.

One of the features of our approach is that the actual output of the protocol is a particular coherent encoding of the target probability distribution. The classical probability distribution can then be recovered by a measurement of the output state. Having such a quantum output is desirable if our protocol is to be embedded in a larger algorithm where the preparation is just an initial step. Examples where this is assumed include hitting algorithms \cite{2010_Krovi,2011_Magniez_SIAM}, and algorithms which require where from a (renormalized) part of the distribution \cite{2010_Krovi, 2014_Paparo}. We point out that this property is not a necessary feature of all quantum algorithms for mixing -- there are promising approaches which utilize decoherence to speed up mixing \cite{2007_Richter}, which may preclude a coherent output.
The property that the output is a coherent encoding of the target distribution is also maintained in extensions of our protocol, which we describe in the next section.
 \section{Lower bounds and extensions}
 \label{sect5}
 The approach we have described in the previous section trivially extends to monotonically increasing distributions as well -- since the trace distance is invariant under the simultaneous  permutations of the elements of the inputs, the same proof holds, where we use "ladder distributions" which are reversed in the order of the probabilities.
 However,  the approach can be further extended to strongly unimodal distributions, and beyond, if additional knowledge about the target stationary distribution is assumed.
 At the end of this section, we will explain how such extensions can be obtained.
 Before this we will address two natural theoretical questions which arise from our approach.
 
 First, in the previous section we have only provided an upper bound of the distance of the set $S$ and an unknown monotonically decaying distribution. A-priori, it is not inconceivable that, for the restricted case of monotonically decaying distributions, it may be possible that there exists a significantly better choice, with a better bound -- perhaps achieving a constant distance, instead of a $\log(N)$ dependance. Here we will show that a significant improvement of our result is not possible.

Second, in our setting we have assumed a specific type of prior knowledge of the target stationary distribution. It is a fair question whether such  knowledge, along with the capacity to prepare particular initial distributions, may already offer a significant speed up in the case of classical mixing. If this were the case, our result should not be considered as a true speed up of classical mixing.  However, we show that the type of assumption we impose for the quantum algorithm does not help in the classical approach. 

\subsection{Lower bounds}

The cornerstone of our result relies on the fact that
there exists a $\log(N)$-sized set of distributions in $D^{\geq}_N$, which is no more than $\log^{2}(N)$ far (in terms of fidelity) from any distribution $\pi$ in $D^{\geq}_N$.
It is a fair question whether the $\log(N)$ dependence can be dropped altogether, and be replaced by a constant, in the complexity of the mixing algorithm.
A necessary precondition for this, in the case of our approach, is the following claim:
\\

\noindent \textbf{Claim 1} There exists a constant $0 \leq \eta < 1$, and a family of (arbitrary) distributions $\{ \mu^{(N)} \}_{N}$, one for each state space size $N$, such that for every $N \in \mathbbmss{N},$ and for every $\pi \in D^{\geq}_N$ we have that $D(\mu^{(N)}, \pi) \leq \eta$.
\\

If \textbf{Claim 1} were to be true, and if the coherent encodings of distributions $\mu^{(N)}$ were efficiently constructable (say in time $O(\textup{polylog}(N))$), then this would constitute a significant improvement over our result. 
To get a bit of intuition, consider a generalization of Claim 1, where $\pi^k$ are arbitrary distributions. In this case the claim clearly does not hold. Consider any family $\{ \mu^{(N)} \}_N$. Then for $N\in \mathbbmss{N},$ let $\mu_{min} = \min_j (\mu^{(N)})_j$ be the smallest probability occurring in $\nu^{(N)}$, and let $j_{min} = \textup{argmin}_j (\mu^{(N)})_j$ be the position of the smallest probability. Then we can choose $\pi$ to be the Kronecker delta distribution at position $j_{min}$ which yields the distance $1-\mu_{min} \geq 1- 1/N,$ which converges to 1 with the state space size $N$. 

Unfortunately, similar simple arguments cannot be straightforwardly utilized in the case when $\pi$ is in $D^{\geq}_N$, and the proof is a bit more involved. Our theorem (stated later) is the negation of Claim 1, and we prove it by contradiction.

First, we show that \textbf{Claim 1} implies a seemingly stronger claim denoted \textbf{Claim 2}:\\

\noindent \textbf{Claim 2} There exists a constant $0 \leq \eta < 1$, and a family of distributions $\{ \mu^{(N)} \}_i$, $\mu^{(N)} \in D^{\geq}_N$ defined for any state space size $N$, such that for every $N \in \mathbbmss{N},$ and for every $\pi \in D^{\geq}_N$ we have that $D(\mu^{(N)}, \pi) \leq \eta$.
\\

The difference between Claim 1 and Claim 2 is that in Claim 2, the family $\mu^{(N)}$ is in the sets of monotonically decaying distributions.
We prove this generically by sorting.
Let $\textup{Sort}$ be a map from the set of distributions to the set of decaying distributions, which, for any distribution $\pi$ outputs a distribution $\pi'$ where the entries of $\pi'$ are sorted in a decreasing fashion. While the sorted distributions may not be unique, we may, assume $\textup{Sort}$ picks a unique sorting which preserves the original ordering in the case multiple positions (states) have the same probability.
Then the following lemma holds:
\LE
Let $\mu$ be an arbitrary $N-$state distribution and let $\pi \in D^{\geq}_N$ be a monotonically decaying distribution. Then $D(\pi, \mu) \geq D(\pi, \textup{Sort}(\mu))$.
\EL
\proof
We will prove that switching any two (adjacent) elements $i,j$ in the distribution $\mu$, which do not obey $\mu_i \geq \mu_j$ can only decrease the distance from $\pi$. 
This suffices for our lemma, as iterating such switching upon the distribution (in an, for us unimportant order) $\mu$ constitutes the well-known \textit{Bubble\ sort} algorithm for sorting, which converges to a sorted distribution (list) in at most $N^2$ steps. Since each step only decreases the distance, by transitivity we have our claim.

We prove this by direct verification. Let $i,$ $j=i+1$ be such that $\mu_{i} \leq \mu_j$, and let $\mu'$ be the distribution obtained from $\mu$ by switching labels $i$ and $j$.
Then we have: 
\EQ{
D(\mu, \pi) = \sum_{k \not= i \ \textup{and}\ k\not= j} 1/2 | \mu_k - \pi_k| + 1/2 (|\mu_i - \pi_i| + |\mu_j - \pi_j|)\\
D(\mu', \pi) = \sum_{k \not= i \ \textup{and}\ k\not= j} 1/2 | \mu_k - \pi_k| + 1/2 (|\mu_j - \pi_i| + |\mu_i - \pi_j|),
}
so
\EQ{
2 (D(\mu, \pi) - D(\mu', \pi)) &=  |\mu_i - \pi_i| + |\mu_j  - \pi_j| - ( |\mu_j - \pi_i| + |\mu_i - \pi_j|) \\ &=|x| + |x + (\epsilon + \delta)| - |x+\epsilon| - |x+ \delta|
}
 where $x = \mu_i - \pi_i, \epsilon = \mu_j - \mu_i \geq 0$ and $ \delta = \pi_i - \pi_j \geq 0$. We can without the loss of generality assume $\epsilon \leq \delta$.
 We have five possibilities: a) $x \geq 0$; b) $ 0 \geq x \geq - \epsilon$, c) $-\epsilon  \geq x \geq -\delta$; d) $-\epsilon - \delta \leq x \leq - \delta,$ and e) $x< -\delta -\epsilon$. In the case $a)$ the difference is zero, so the claim holds. In the case $b)$ we have 
 \EQ{
 -x + x + \epsilon + \delta -x - \epsilon -x -\delta =-2x \geq 0.
 } 
 In the case $c)$ we have: 
 \EQ{
 -x + x + \epsilon + \delta + x +\epsilon -x - \delta = 2 \epsilon \geq 0.
 }
In the case $d)$ we get $-x+x + \epsilon + \delta +x + \epsilon + x + \delta = 2 (\epsilon+ \delta) + 2x \geq 0 $. 

Finally, in e) we have $-x -x - \epsilon -\delta + x + \epsilon + x + \delta = 0.$ This proves the lemma. \qed

Thus, for our main goal, it will suffice to show that \textbf{Claim 2} does not hold.

Next, note that  \textbf{Claim 2} also implies the claim that the distributions $\nu$ are at most constantly far from all "ladder" distributions $\{ \sigma^i \}_i$ defined over the same state space.
Moreover, by the extremity of these distributions, the inverse holds as well - $\nu$ being at most $\eta$ far from all the  $\{ \sigma^i \}_i$ distributions also implies that $\mu$ is at most $\eta$ far from any monotonically decaying distribution. However, we shall not be using the inverse claim.

In the remainder we will assume $\mu$ is a decaying distribution over $N$ states, and also that $\{ \sigma^i \}_i$ are the "ladder" distributions we have described in Eq. (\ref{ladders}).

Recall that $\mu$ (since it is in $D^{\geq}_{N}$) can be represented in the convex-linear basis of $\{ \sigma^i \}_i$ (c.f. Eq (\ref{eq1})): 
\EQ{
\mu = \sum_{i=1}^{N}  q_i \sigma^i,
}
where for $1 \leq i < N$ $q_i = i  (\pi_i - \pi_{i+1})$ and $q_N = N \pi_N$. As we have seen, as $\mu$ is decaying, we have that $q = (q_i)_i$ is a probability distribution as well, uniquely specified for every $\mu$.
Retracing the steps of Theorem \ref{first}, we have also defined vectors $\{ v_k \}_{k=1}^{N}$ with $(v_k)_i = \dfrac{\min\{i,k \}}{\max\{i,k \}}$ using which we can express the trace distance between $\mu$ and $\sigma^k$ as:
\EQ{
D(\mu, \sigma^k) = 1- v_k^T q.
}
Recall, the matrix $V$ collects the vectors $v_k$ as rows (and columns, since it is symmetric). Then the $k^{th}$ row of the vector $Vq$ equals $1 - D(\mu, \sigma^k)$.
The claim we seek to show is that $ \max_{q} \min_{k}  v_k^{T}q$, that is the maximal overlap of the $v_k$ vectors, optimized over all probability distributions $q$ depends on $N$ and decays to zero in the limit of infinite state space.
To establish this, we will use the specific properties of the matrix $V$ and a few simple results from the theory of convex spaces.
First, to remind the reader, the matrix $V$ is defined as $V_{ij} = \dfrac{\min\{i,j \}}{\max\{i,j \}}$, thus it is a symmetric matrix. Moreover, as we will show later, it is also also invertible. In the language of convex spaces this simply means that a point in a convex set is uniquely specified by the distance of that point from the extreme points of the convex set.
Next, intuitively, the minimum of distances from the points $v_k$ (understood as points in an $N-$dimensional Euclidian space) is attained by a point in the convex hull of those points, which is equally far from all of them. However, the validity of this claim can depend on the choice of the distance measure. We shall thus prove this claim for our case. For the remainder of the proof, we will have to evaluate the distance attained, that is, the value of  $\max_{q} \min_{k}  V.q$.
For this, we will use the explicit inverse of the matrix $V$, which we give as a separate lemma for clarity. 
\LE
Let $N \in \mathbbmss{N}$, and let $V$ be an $N \times N$ matrix defined with $V_{ij} = \min\{i,j\}/\max\{i,j \}$. Then $V^{-1}$ exists, and it is a symmetric, diagonally strictly dominant, tri-diagonal matrix specified with:
\EQ{
\left[V^{-1}\right]_{i,i+1} = -\dfrac{i(i+1)}{2i+1};\\
\textup{for} \ i<N\ ,\left[V^{-1}\right]_{i,i} = \dfrac{4i^3}{(2i-1)(2i+1)},  \textup{and}\\
\left[V^{-1}\right]_{N,N} = \dfrac{N^2}{2N-1}.}
\EL
\proof To see that $V$ multiplied with $V^{-1},$ as defined above, is the identity can be easily checked by examining the diagonal and non-diagonal elements of $VV^{-1}$ separately, so we omit this here.
What remains to be seen is that $V^{-1}$ is diagonally dominant. For the column (also row, since it is symmetric) $i=1$ it is obvious.  
For $1<i<N$ we have that:
\EQ{
\dfrac{4i^3}{(2i-1)(2i+1)} > \dfrac{i(i+1)}{2i+1}+\dfrac{i(i-1)}{2i-1}\\
\dfrac{4i^3}{(2i-1)(2i+1)} > \dfrac{i(i+1)(2i-1) + i(i-1)(2i+1)}{(2i+1)(2i-1)}\\
4i^2 >2 (2i^2 -1) \Leftrightarrow 2i^2 > 2i^2 -1
}
so the the inequality is strict.
Finally, for $i=N$, we have that 
\EQ{
 \dfrac{N^2}{2N-1} > \dfrac{N(N-1)}{2N-1},
}
which is also a strict inequality. Thus the lemma holds. \qed

We now claim that since $V^{-1}$ is strictly diagonally dominant the optimal value $\max_{q} \min_{k}  v_k ^{T}q$ is attained at a $q$ for which $Vq = \alpha(1, \ldots, 1), \alpha \in \mathbbmss{R}^+$ - in this case, the distribution encoded by $q$ is equally far from all ladder distributions.

We prove this claim by contradiction. The negation of this claim implies that there exists a probability distribution $q'$ such that $Vq' \geq Vq$ (that is, the distribution encoded by $q'$ is closer to all ladder distributions), where the inequality is taken element - wise.
Let $V(q'-q) = y$ with $y=(y_1, \ldots, y_N)^{T}$. Note, $y$ is by assumption a non-negative vector, with at least one positive entry.
Then $(q-q') = V^{-1}y$. Note that since $q$ and $q'$ are distributions, the sum of all elements of the vector $(q-q')$ is $0$. Then, by multiplying both sides of the last equality with the row $(1,\ldots,1)$ from the left
we get 
\EQ{0 = (1,\ldots,1)V^{-1}y .\label{eq6}}

Since $V^{-1}$ is strictly diagonally dominant (both column and row-wise since it is symmetric), the row $ (1,\ldots,1) V^{-1}$ is a strictly positive row vector.
But then the inner product $ (1,\ldots,1)V^{-1}y$ must be strictly larger than zero, which is a contraction with Eq. (\ref{eq6}).

Thus $q$ such that $Vq = \alpha(1, \ldots, 1)^{T}$ maximizes the minimal overlap (inner product between rows of $V$ and $q$).
Finally, we evaluate $\alpha$.
By applying $V^{-1}$ from the left from the last equality, and multiply both sides with the row $(1,\ldots,1)$ from the left, since $q$ is a probability vector, we get:
\EQ{
1 = \alpha (1,\ldots,1)V^{-1}(1, \ldots, 1)^{T}.
}
In other words, $\alpha$ is the inverse of the sum of all the elements in $V^{-1}$. We then have that:
\EQ{
1/\alpha = \left(\sum_{i=1}^{N-1} \dfrac{4i^3}{(2i-1)(2i+1)} \right) + \dfrac{N^2}{2N-1} - 2\sum_{i=1}^{N-1} \dfrac{i(i+1)}{2i+1}
}
The expression above can be much simplified.
We have that:
\EQ{
1/\alpha = 2 \sum_{i=1}^{N-1}\left( \dfrac{2i^3}{(2i-1)(2i+1)} - \dfrac{i(i+1)}{2i+1}\right) + \dfrac{N^2}{2N-1}   \\
= -2 \sum_{i=1}^{N-1}\left( \dfrac{i^2-i}{(2i-1)(2i+1)}\right) + \dfrac{N^2}{2N-1}  \\
=  -2\left( \dfrac{N(N-1)}{4N-2} -  \sum_{i=1}^{N-1} \dfrac{i}{(2i-1)(2i+1)}\right) + \dfrac{N^2}{2N-1}  \\
=   \sum_{i=1}^{N-1} \dfrac{2i-1}{(2i-1)(2i+1)} + \dfrac{N-1}{2N-1}  + \dfrac{N}{2N-1}   \\
  =  \sum_{i=1}^{N-1} \dfrac{1}{(2i+1)} +1 =   \sum_{i=1}^{N-1} \left( \dfrac{1}{(2i+1)}+ \dfrac{1}{(2i)} \right) - \sum_{i=1}^{N-1} \dfrac{1}{(2i)}+1 =  H_{2N}  - \dfrac{1}{2N} - \dfrac{1}{2}H_{N} +\dfrac{1}{2N}   \\
    = H_{2N} - \dfrac{1}{2}H_{N} = \dfrac{1}{2} H_{N-1/2} + \log(2),
 }

where $H_x$ is the (generalized) Harmonic number function, for $x>0$ defined with
\EQ{
H_x =  \sum_{k=1}^{\infty}\left( \dfrac{1}{k} - \dfrac{1}{k+x} \right).
}

Thus we have that 
\EQ{
1/\alpha > \dfrac{1}{2} H_{N-1/2} +\log(2).
}

Harmonic numbers and the natural logarithm have a well-understood relationship, and for our purposes it will suffice that $H_{N-1/2} \geq \log(N) $.
It follows that
\EQ{
1/\alpha > \dfrac{1}{2} \log(N)\\
\alpha < \dfrac{2}{\log(N)}.
}
The last inequality immediately implies our claim as, putting all the observations together, we then have that:
\EQ{
\min_{\sigma \in D_{N}} \max_{\pi \in D^{\geq}_{N}}  D(\pi,\sigma) > 1 - \dfrac{2}{\log(N)},
}
where $D_{N}$ denotes the set of all $N-$state distributions.
That is, the optimal family of distributions, one for each state space size, will be at least $1 - \dfrac{2}{\log(N)}$ far from the worst case distance from a distribution in $D^{\geq}_N$.
This shows not only that the \textbf{Claim 1} does not hold, but that our algorithm is not far from optimal, and that if one had access to an oracle (or an efficient construction), of the coherent encodings optimal distributions, the overall algorithm would have the efficiency depending on $\log^{1/2}(N).$ In contrast, we constructively achieve $\log^{3/2}(N).$ While the first scaling is clearly better, the difference is moderate, given that both are polylog.
For completeness, we phrase the results above as a theorem.
\TH
For any family of distributions $\{ \mu^{(N)} \}_{N \in \mathbbmss{N}},$ each defined over the state space of size $1<N \in \mathbbmss{N}$, and for every $N,$ there exists a monotonically decaying distribution $\pi \in D^{\geq}_N,$ such that
\EQ{
D(\pi, \mu^{(N)}) > 1- \dfrac{2}{\log(N)}.
} 
\HT

We end this subsection with a remark on the $V$ matrices, as a curiosity. We have in the proof above, shown that the inverse of a $V$ matrix is, technically speaking, an inverse of a strictly diagonally dominant Stieltjes matrix, hence also an M-matrix. The problem of characterizing non-negative matrices, whose inverse is an M-matrix has attained significant interest in the field of matrix analysis \cite{1995_McDonald}. One result presented in that research is a characterization of so-called generalized ultrametric matrices, the subset of which is shown to, by inversion, yield diagonally dominant Stieltjes matrix, as is $V^{-1}$ in our case. It is curious to note that, however, the $V$ matrices we use do not fall into the class of the characterized generalized ultrametric matrices, thus may have independent interest in the field of matrix analysis.

\subsection{Classical mixing for decaying distributions}

The results of Section \ref{Sect3} show that, in the case of quantum mixing through un-searching, the overall mixing time strongly depends on the initial state - the mixing time is proportional to the inverse of the square-root of the fidelity between the initial state and the targeted stationary distribution state. A similar statement holds when we consider the trace distances between the initial distribution (encoded by the quantum state) and the stationary distribution. This is clear as the trace distance (of classical distributions), and fidelity (of their coherent encodings) are tightly connected.
Moreover, this dependence of the mixing time on the distance between the initial and target state is robust - regardless of what particular initial state we pick, the mixing time just depends on the distance.

In the classical case, intuitively it is also clear that starting from a distribution, which is close to the target distribution, must speed up mixing. As an extreme example, if we wish to achieve mixing within $\epsilon$, and we are given an initial state which is already within $\epsilon$ from the target, the mixing time (in the sense of the number of required applications of the walk operator) is zero.
The question is, is the improvement as robust in the classical case, as it is in the quantum? Here we show that in the classical case, it is not, and being close to the target distribution helps just moderately.
To show this we first clarify a fact about classical mixing times, the definition of which we repeat for the benefit of the reader.
The mixing time $\tau(\epsilon),$ within error $\epsilon$, for Markov chain $P,$ with stationary distribution $\pi$ is defined as:
\EQ{
\tau(\epsilon)  = \min \left\{t \vert D(P^{t} \sigma,\pi)\leq \epsilon, \forall \sigma \right\}
}
where $D(\pi, \sigma)$ denotes the total variational distance on distributions $\pi, \sigma$, so $D(\pi, \sigma) = 1/2 \sum_j |\pi_j  - \sigma_j|$.
The mixing time asks that the state $P^{t} \sigma$ be $\epsilon$ close to $\pi$ for all initial states $\sigma$, that is, it looks for the worst case initial $\sigma$. By convexity, and the triangle inequality, the worst case initial state $\sigma$ will be a Kronecker-delta distribution with total mass at some state space element.
 
We can  introduce an analogous mixing time quantity, \emph{relative mixing}, which extends standard mixing time in that the initial state is guaranteed to be within $\eta$ from the target state:
\EQ{
\tau_{\eta}(\epsilon)  = \min \left\{t \vert D(P^{t} \sigma,\pi)\leq \epsilon, \forall \sigma\  s.t.\ D(\sigma,\pi)\leq \eta  \right\}.
}
Now, suppose we are given a Markov chain $P$, and we wish to evaluate a bound on $\tau_{\eta}(\epsilon)$ for this Markov chain.
In order to capture robust properties, the definition above asks for the worst case as well (as the distance requirement it must hold for \emph{all} $\sigma\  s.t.\ D(\sigma,\pi)\leq \eta$), so to bound the relative mixing times, we can construct the following distribution $\rho:$\
\EQ{
\rho =(1-\eta) \pi + \eta{\sigma_{worse}},
}
where we choose $\sigma_{worse}$ to be the worst-case initial state for MC $P$ if we wish to mix it within $\epsilon/\eta$.
Then we have that:
\EQ{
\dfrac{1}{2}||(1-\eta) \pi + \eta \sigma_{worse} - \pi || = \eta \dfrac{1}{2}|| \sigma_{worse} - \pi || \leq \eta,
}
so $\rho$ is within $\eta$ distance from $\pi,$ as required.
Now, we are looking for an integer $t \geq 0,$ such that:
\EQ{
\dfrac{1}{2}||P^t \rho  - \pi || \leq \epsilon.
}
Then we have:
\EQ{
\dfrac{1}{2}||P^t \rho  - \pi || = \dfrac{1}{2}||(1-\eta)P^t \pi + \eta P^t \sigma_{worse} - \pi || =  \dfrac{1}{2}||(1-\eta) \pi + \eta P^t \sigma_{worse} - \pi || = \eta D(P^t \sigma_{worse}, \pi),
}
hence, we require a $t$ such that
\EQ{
D(P^t \sigma_{worse}, \pi) \leq \dfrac{\epsilon}{\eta}.
}
However, since $\sigma_{worse}$ was chosen to be the worst case state for mixing within $\dfrac{\epsilon}{\eta}$, we have that
\EQ{
\tau_{\eta}(\epsilon)  \geq \tau(\epsilon/\eta).
}
Thus the lower bound of the relative mixing time is just the standard mixing time, where $\epsilon$ is replaced with $\epsilon/\eta.$ Thus, we get the following lower bounds for relative mixing:
\EQ{
\lambda_2 \dfrac{1}{\delta} \log\dfrac{1}{2 (\epsilon/\eta)} \leq \tau(\epsilon/\eta) \leq  \tau_{\eta}(\epsilon).}
It is now clear that the relative mixing has the same dependence on $1/\delta$, hence the improvement is slight. To make a fair comparison to the quantum mixing case we have shown, we can set $\eta = 1/2$ (for our algorithm, the trace distance is always larger than this), and 
see that the lower bound for classical mixing is lower bounded by $O(1/\delta \log(1/(4 \epsilon))),$ which is essentially the same scaling, as for standard mixing time.

For completeness, we point out that there have been prior works asking a related question, which establish that the mixing time is an essentially robust quantity, independent from the setting (that is, in what context) the mixing is applied. We refer the reader to \cite{1997_Aldous} for a collection of such results.

\subsection{Extensions}
While the main theorem we have used in our approach assumes monotonic (decaying or increasing) distributions, this can be easily further extended.
Assume, for instance, we know that $\pi$, the target distribution over $N=2^n$ elements only decays to some element $k,$ its behavior is unknown from that point on, and the total support up to element $k$ is $p=\sum_{i=1}^{k} \pi_i$. 
 Consider for the moment the truncated distribution $\tilde{\pi},$ obtained by setting all probabilities after $k$ to zero and re-normalizing (by multiplying with $1/p$.)
By Theorem $\ref{first},$ we know that there exist a (efficiently constructable) log-sized set  $S$ of "ladder" distributions over $N,$ such that for at least one of them, $\sigma$, it holds that
$D(\sigma, \tilde{\pi}) \leq 1 - (n+1)^{-1}/2.$
But then we have, by the triangle inequality, homogeneity of the trace norm, and the fact that the maximal distance is unity, that:
\EQ{
D(\sigma, \pi) \leq p D(\sigma, \tilde{\pi}) + (1-p) = 1 - \dfrac{p}{2(n+1)}.
}
This implies that the total complexity of the mixing algorithm we have described, applied to this setting will be increased, multiplicatively, by $p^{-1}$.
Note that the same reasoning will hold, in the mirrored case, where we know that $\pi$ is increasing from some element $k,$ with corresponding support of $p$.

This simple observation already allows us to efficiently prepare target distributions whose probability mass functions are convex (decaying to some element, and increasing from that element). To see this note that either the mass of the distribution prior its minimum, or after, must be above or equal to $1/2$. Thus, we can simply run the algorithm assuming both options which then yields just a constant multiplicative overhead of 4 (two runs $\times (1/2)^{-1}$).

Another extension of this observation is the case where relative to a known order,  the distribution is, for a known contiguous subset of the state space elements (with total weight $p$), decaying or increasing. In this case as well, the mixing time only suffers a $1/p$ pre-factor. In particular, this implies that distributions which are strictly unimodal (meaning increasing to some element, and decreasing from that element) can also be efficiently prepared, provided the mode $k$ is known. To see this holds, note that in the strictly unimodal case as described, the total mass of the probability distribution either up to the $k^{th}$ element, or after, must be above $1/2$. Then, the ladder distributions would only be constructed to the $k^{th}$ element, which again can be done with $\textup{polylog}(N)$ overhead.
Unfortunately, for this case, the knowledge of the position of the mode $k$ is neccessary -- strictly unimodal distribution also contain the Kronecker-delta distribution. For our approach the capacity to efficiently mix to (a distribution arbitrarily close to) an arbitrary Kronecker-delta distribution would immediately imply efficient mixing for all distributions. This is beyond what we can claim.

\section{Discussion}
 \label{Sect6}
 In this work, we have addressed the problem of attaining stationary distributions of Markov chains using a quantum approach. We have built on observations, originally made by  Richter and Childs, that quantum \emph{hitting} algorithms run \emph{in reverse} can serve as mixing algorithms. These observations initially received little attention due to their apparent inefficiency -- an a-priori square-root scaling with the system size $N$.
 We have shown, in contrast, that in the cases when it is beforehand known that the target distribution is decaying, relative to a known order on the state space, the dependency on the system size is only $\log^{3/2}(N)$. We have also shown the essential optimality of this bound for our approach, in particular, that an explicit dependence on the system size is unavoidable and logarithmic.
 Following this we have shown how our approach easily extends to a much wider class of distributions, including concave distributions, but also strictly unimodal distributions, where the position of the mode is known.
 Unfortunately, it is the case that such assumptions are often not satisfied in many physics-inspired applications which require mixing of Markov chains. For instance, in statistical physics, the mode is often the quantity explicitly sought, when the distribution is known to be unimodal. In other uses, e. g. the computation of a permanent of the matrix, the underlying state space is not simply characterized at all, and knowing the order would already imply the solution to the problem.
 
 Nonetheless, other applications involving Markov chain mixing, such as artificial intelligence  \cite{2014_Paparo} and applications relying on bayesian inference (which often rely on MCMC) may have more instances where our approach may yield a genuine quantum speed-up. Moreover, the quantum algorithm we have provided realizes a coherent encoding of the stationary distribution, which can be used as a fully quantum subroutine, for instance in the preparation of initial states in e.g. hitting algorithms \cite{2010_Krovi,2011_Magniez_SIAM}.
 
Another possibility includes settings where the shape of the target distribution is known (say Gaussian), and the mode is also,  however, we are interested in learning higher moments of the distribution by mixing and sampling. For instance, all correlations of Gaussian states in quantum optics are captured by the second moments, whereas the mode and mean coincide, and reveal nothing about correlations. We leave the applications of our results for future work.
From a more theoretical point of view, the results of this work highlight another difference between classical and quantum mixing, in particular, the approaches which rely on reversing hitting algorithms. In the classical mixing case, the choice of the initial state does not substantially contribute to overall mixing efficiency. In contrast, in the quantum case, improvements in the choice of the initial state can, as we have shown, radically alter the overall performance.

While the conjecture that quantum approaches to mixing can yield a generic quadratic speed-up in all cases remains open, our approach extends the class of Markov chains for which such a speed up is possible. Notably, unlike in other studied cases where speed up has been shown, our assumptions lay only on the structure of the stationary distribution of the Markov chain, rather directly on the structure (underlying digraph) of the Markov chain itself.

 \noindent\textbf{Acknowledgments:\\}
The authors acknowledge the support by the Austrian Science Fund (FWF) through 
the SFB FoQuS F 4012, and the Templeton World Charity Foundation grant TWCF0078/AB46. VD thanks Markus Heyl  and Nicolai Friis for many insightful discussions.

%\bibliography{biblio.bib}{}
%\bibliographystyle{unsrt}

\bibliographystyle{unsrtnat}

\end{document}